\documentclass[11pt]{article}
\usepackage{graphicx,amssymb,amsfonts,amsthm,amsmath,amssymb,amscd,amstext,verbatim}

\def\be{\begin{equation}}
\def\ee{\end{equation}}

\def\csch{\operatorname{csch}}

\def\tr{\operatorname{tr}}

\begin{document}

\begin{flushright}
YITP-SB-12-34
\end{flushright}

\begin{center}
\vspace{1cm} { \LARGE {\bf Tracing Through Scalar Entanglement}}

\vspace{1.1cm}

Christopher P. Herzog and Michael Spillane

\vspace{0.8cm}

{\it C.~N.~Yang Institute for Theoretical Physics \\ Stony Brook University, Stony Brook, NY  11794}

\vspace{0.8cm}

\end{center}

\begin{abstract}
\noindent
As a toy model of a gapped system, 
we investigate the entanglement entropy of a massive scalar field in 1+1 dimensions at nonzero temperature.
In a small mass $m$ and temperature $T$ limit, 
we put upper and lower bounds on the two largest eigenvalues of the covariance matrix used to compute the entanglement entropy.  We argue that the entanglement entropy has $e^{-m/T}$ scaling in the limit $T \ll m$.
We comment on the relation between our work and the Ryu-Takayanagi proposal for computing the entanglement entropy holographically.
\end{abstract}

\pagebreak
\setcounter{page}{1}

\section{Introduction}

The notion of entanglement entropy (and more generally quantum entanglement) looms large in theoretical physics today.  
Entanglement entropy may be a good order parameter for topological phase transitions in condensed matter systems.  For conformal field theories in 1+1 dimensions, numerical computation of the entanglement entropy provides a rapid way to calculate the central charge $c$.  In relativistic field theories more generally,
certain special kinds of entanglement entropy show monotonicity properties under renormalization group flow
 \cite{Casini:2006es,Casini:2012ei}.  See \cite{Calabrese:2009qy} and \cite{EislerPeschel} for reviews.
 

To compute the entanglement entropy for a quantum mechanical system, we must first divide the associated Hilbert space up into two pieces.  Usually, the division is made with respect to spatial regions $A$ and complement $\bar A = B$. 
We find the reduced density matrix $\rho_A \equiv \tr_{B} \rho$ by tracing over the degrees of freedom in $B$.  Finally, the entanglement entropy is defined to be
\be
S \equiv - \tr \rho_A \log \rho_A \ .
\ee

It is surprising that even for what many consider to be the simplest field theoretic system -- a massive scalar field in 1+1 dimensions -- the entanglement entropy has thus far been computed analytically only in certain limits.
In the limit $m=0$, one can use results from conformal field theory \cite{Calabrese:2004eu,Korepin:2004zz}.
In particular, for the massless scalar field on the cylinder $\mathbb{R} \times S^1$ where ${\mathbb R}$ is interpreted as the time direction, one has
\be
\label{circleSE}
S = \frac{1}{3} \log \left( \frac{L}{\pi \epsilon} \sin \frac{\pi \ell}{L} \right) + c_0 \ ,
\ee
where $L$ is the circumference of the $S^1$, $\ell$ is the length of the interval, $\epsilon$ is a UV regulator and $c_0$ is a constant that depends on the regulation scheme.  (In fact, for the massless scalar, there is an additional IR divergence,  and $c_0$ depends also on an IR cutoff.)
Reinterpreting $S^1$ as a Euclidean time direction, one obtains a result at nonzero temperature $T=1/\beta$ for the scalar on $\mathbb{R}$.
\be
\label{thermalSE}
S =  \frac{1}{3} \log \left( \frac{\beta}{\pi \epsilon} \sinh \frac{\pi \ell}{\beta} \right) + c_0 \ .
\ee
When $m \neq 0$ for the scalar field on $\mathbb{R}^2$, Huerta and Casini \cite{Casini:2005zv} 
  have shown that the entanglement entropy can be computed from the solution to a certain Painlev\'{e} equation.  Their work allows analytic access to the small and large mass limits.
For $m\ell \ll 1$, one obtains
\be
S \sim \frac{1}{3} \log \frac{\ell}{\epsilon} + \frac{1}{2} \log\left( \frac{\log(m \epsilon)}{\log (m \ell)} \right) \ ,
\ee
while for $m\ell \gg 1$, one finds instead exponential suppression\footnote{%
 A generalization was obtained by Doyon and collaborators \cite{Doyon:2008} and \cite{Doyon2} allowing for multiple masses.}
\be
S \sim \frac{1}{16} \sqrt{\frac{\pi}{m \ell}} \, e^{-2 m \ell} \ .
\ee

Ideally, one would like to understand the case where $m$, $T$, and $1/L$ are all nonzero.
Numerically, the entanglement entropy can be computed with ease using a generalization \cite{Peschel} of the procedure introduced by Srednicki \cite{Srednicki}.  One realizes the scalar field as the continuum limit of an $N$-site harmonic chain.  
For such a chain, one introduces two point functions $\langle \phi_i \phi_j \rangle$ and $\langle \pi_i \pi_j \rangle$ of the oscillator positions and conjugate momenta respectively.  Restricting now to an interval $n \epsilon = \ell<L$ where $1 \leq i,j, k \leq n$, one constructs the $n \times n$ matrix 
\be
\label{defC2}
(C^2)_{ij} \equiv \sum_{k=1}^n \langle \phi_i \phi_k \rangle \langle \pi_k \pi_j \rangle \ .
\ee
The entanglement entropy is then
\be
\label{SEfromC2}
S = \tr \left[ (C +1/2) \ln (C+1/2) - (C-1/2) \ln (C-1/2) \right] \ .
\ee
To our knowledge, this quantity has not been computed analytically for the real scalar field with two or more of the quantities $m$, $T$, and $1/L$ nonzero.  Happily, with today's desktop computers, it is relatively quick to diagonalize $C$ numerically for $N \sim 10^3$.  Ref.\ \cite{BoteroReznik} provides a numerical analysis of the harmonic chain using this approach.

In this paper, we take some steps toward an analytic understanding of the eigenvalues of $C$.  
As noted in \cite{BoteroReznik}, the parity operator $P$ commutes with $C$ where parity here is a reflection of the circle $S^1$ with respect to the midpoint of the interval.  Thus, one may divide $C$ into even and odd parity blocks $C_e$ and $C_o$.  
We compute the two partial traces $\tr C_e^2$ and $\tr C_o^2$ in the limit $m, T \ll 1/L$.  As the spectrum of $C^2$ is bounded below by 1/4, these traces give us upper bounds on the two largest eigenvalues of $C$.  A variational approach gives a lower bound to the largest (parity even) eigenvalue.  These bounds in turn give us some intuition for the $m$, $T$, and $L$ dependence of the entanglement entropy in the limit $m,T \ll 1/L$.  

The original motivation for this project came from our interest in the Ryu-Takayanagi proposal \cite{Ryu:2006bv} for computing the entanglement entropy of field theories with dual holographic classical gravity descriptions.  
Given two complementary regions $A$ and $B$ in the field theory, the Ryu-Takayanagi proposal associates a nonzero
 $S_A - S_B$ to gravity descriptions with black holes, while in the absence of such defects $S_A = S_B$.  
 In the dual field theory, the existence of a black hole typically implies deconfined gauge theory degrees of freedom \cite{Witten:1998zw,Herzog:2006ra}. 
 
We may contrast this result with the quantum mechanical point of view where at $T=0$, the density matrix is constructed from a pure state. (We are assuming the existence of a unique ground state.) 
  It follows from a Schmidt decomposition of the Hilbert space that for pure states $S_A = S_B$ (see for example \cite{EislerPeschel}).
  However, at any nonzero temperature, regardless of the presence of deconfined degrees of freedom, the density matrix is not constructed from a pure state and one would generically expect $S_A \neq S_B$.  As gauge theories are  more difficult to study than the free scalar field and as the entanglement entropy of the free scalar field has not yet been completely understood, our toy model of confinement in this paper is  a 1+1 dimensional massive scalar field on a circle at $T>0$.  Morally, the regime $T \ll m$ can be thought of as ``confining''.\footnote{%
 Klebanov et.~al.\ \cite{Klebanov:2007} were the first to consider the entanglement entropy of confining theories from a holographic perspective.  Their work at zero temperature was later followed up by lattice computations \cite{Buividovich, Velytsky, Nakagawa:2011su}.}   One of our results is that in this regime, the entanglement entropy difference does not vanish but rather scales 
  as\footnote{%
 After finishing this work, we became aware of ref.\ \cite{Cacciatori:2008qs} where the same exponential behavior was found for a ``renormalized thermal entropy'' similar in some respects to the entanglement entropy we study here.}
\[
S_A  - S_B \sim e^{-m/T} \ .
\]

\section{From the Harmonic Chain to the Scalar Field}

Consider the Hamiltonian for a real free massive scalar field on a circle of circumference $L$ at $T>0$:
\be
H = \frac{1}{2} \int dx \left[ \pi(x)^2 + (\partial_x \phi(x))^2 + m^2 \phi(x)^2 \right] \ .
\ee
We discretize the circle into $N$ points where $L = N \epsilon$:
\be
H = \frac{1}{2 \epsilon} \sum_{j=1}^N \left[ \pi_j^2 + (\phi_{j+1} - \phi_{j})^2 + m^2 \epsilon^2 \phi_j^2 \right] \ ,
\ee
where $\pi(j \epsilon) = \pi_j/ \epsilon$ but $\phi(j \epsilon) = \phi_j$.
The thermal density matrix can be written in terms of $H$ in the standard way:
\be
\rho = \frac{e^{-H/T}}{ \tr (e^{-H/T})} \ ,
\ee
and expectation values are defined via $\langle X \rangle \equiv \tr (\rho X)$.
A short calculation yields the two point functions of the oscillator positions $\phi_j$ and their conjugate momenta $\pi_j$:
\begin{eqnarray}
\label{phiphi}
\langle \phi_j \phi_k \rangle &=&
\frac{1}{2N} \sum_{a=0}^{N-1} \frac{1}{\epsilon \omega_a} \coth \left( \frac{ \omega_a }{2 T} \right) \cos \left( \frac{2 \pi (j-k)a}{N} \right) \ , \\
\label{pipi}
\langle \pi_j \pi_k \rangle &=& 
\frac{1}{2N} \sum_{a=0}^{N-1} \epsilon \omega_a \coth \left( \frac{ \omega_a }{2 T} \right) \cos \left( \frac{2 \pi (j-k)a}{N} \right) \ ,
\end{eqnarray}
where 
\[
\omega_a^2 = m^2 + \frac{4}{\epsilon^2} \sin^2 \frac{\pi a}{ N} \ .
\]
From eqs.\ (\ref{defC2}) and (\ref{SEfromC2}), we may compute the entanglement entropy from the matrix $C^2 = \langle \pi \pi \rangle \cdot \langle \phi \phi \rangle$ where the two point functions are now restricted to the interval $A$: $-s \leq j,k \leq s$.  In terms of $n$, we have $2s+1=n$.  For simplicity, we choose $n$ to be an odd number.  Any dependence on the parity of $n$ should disappear in the large $N$ limit.

The Hamiltonian $H$ is a set of $N$ coupled harmonic oscillators.  Diagonalizing the Hamiltonian, one finds $H = \sum_a \omega_a b_a^\dagger b_a$ where $[b_a, b_b^\dagger] = \delta_{ab}$.  Surprisingly for a free scalar field, the reduced density matrix $\rho_A \sim e^{-H_A}$ can be written in terms of a similar entanglement Hamiltonian $H_A = \sum_k \epsilon_k b_k^\dagger b_k$ (see for example \cite{EislerPeschel}).  Moreover, there is a one-to-one correspondence between eigenvalues $\lambda_k$ of $C^2$  and the energies $\epsilon_k$:
\be
\lambda_k = \frac{1}{4} \coth^2 \frac{\epsilon_k}{2} \ . 
\ee
As the $\epsilon_k$ are real, we conclude that $\lambda_k \geq 1/4$.

\section{Taking Traces}

For a region $-s \leq k \leq s$, the matrix $C^2$ commutes with the parity operator\footnote{%
 Note that $C^2$ commutes with the parity operator for both odd and even $n$.  
 For example, if we indexed $C^2$ from $1\leq k \leq n$, parity would send $k \to n-k+1$.
 }  
  which sends $k \to -k$.
 Thus, we can decompose $C^2$  into even and odd parity pieces, $C^2 = C_e^2 + C_o^2$.    
The matrices $C_o^2$ and $C_e^2$ are then given by
\begin{eqnarray}
C_e^2 &=& 
\frac{1}{4N^2} \sum_{a,b} \frac{\omega_a}{\omega_b} \coth \left( \frac{\omega_a}{2 T} \right)
\coth \left(\frac{\omega_b}{2 T} \right) \frac{\sin \frac{\pi n(a-b) }{N}}{\sin \frac{\pi (a-b)}{N}} 
\cos \frac{2\pi j a}{N} \cos \frac{2 \pi k b}{N} \ , \\
C_o^2 &=&
 \frac{1}{4N^2} \sum_{a,b} \frac{\omega_a}{\omega_b} \coth \left( \frac{\omega_a}{2 T} \right)
\coth \left(\frac{\omega_b}{2 T} \right) \frac{ \sin \frac{\pi n(a-b) }{N}}{\sin \frac{\pi (a-b)}{N}} 
\sin \frac{2\pi j a}{N} \sin \frac{2 \pi k b}{N} \ ,
\end{eqnarray}
While our main interest is a circle with periodic boundary conditions, the eigenvalues of $C_e$ and $C_o$ also allow us to compute the entanglement entropy for an interval of length $s$ sitting at one end of a strip of length $N/2$.  The matrix $C_o$ gives the two point function of  a strip with Dirichlet boundary conditions, while $C_e$ corresponds to Neumann boundary conditions.

The numerics suggest that for small masses ($m L \ll 1$) and low temperatures ($T L \ll 1$), the matrix $C^2$ has only a handful of eigenvalues which are significantly different from $1/4$.  
The largest of these eigenvalues corresponds to an eigenvector with even parity, while the second largest has odd parity.  We  approximate these eigenvalues by computing $\tr C_e^2$ and $\tr C_o^2$. We find in the even sector that 
\begin{eqnarray}
\label{trCe2}
\tr C_e^2 &=&
 \frac{1}{2\pi m L} \coth \left(\frac{m}{2T} \right) \left[ \gamma + \ln \left( \frac{4N \sin ( \pi r)}{\pi} \right) \right] 
 + \frac{r^2}{4} \csch^2 \left(\frac{m}{2T}\right)
\nonumber \\ 
  && 
   +\frac{1}{4} \Biggl[ s  +
 \frac{11}{12} - \frac{1}{\pi^2} \nonumber \\
 && + \frac{1}{2 \pi^2} \left( -2 + \gamma + 4 \ln \frac{2N}{\pi} - 3 \ln \frac{4 N \sin ( \pi r)}{\pi} \right)
 \left( \gamma + \ln \frac{4N \sin ( \pi r)}{\pi} \right) \Biggr] \nonumber 
\\
 &&
- \frac{3m L}{32 \pi^3} \coth \left( \frac{m}{2 T} \right)
  \left[ \operatorname{Li}_3(e^{2 \pi i r})+ \operatorname{Li}_3(e^{-2 \pi i r}) - 2 \zeta(3) \right]
\nonumber \\
&&
+ O((m L)^2,  e^{-2\pi/ T L}, \log N /N)
 \ ,
\end{eqnarray}
and that in the odd sector
\begin{eqnarray}
\label{trCo2}
\tr C_o^2 &=& 
\frac{1}{4} \Bigl[
s + \frac{1}{12} - \frac{3}{2 \pi^2} +
\frac{1}{2\pi^2}
\left( \gamma-1+ \ln \frac{4N \sin ( \pi r)}{\pi}\right)^2 
\Bigl]
\nonumber \\
&&
+ O((m L)^2,  e^{-2\pi/ T L}, \log N /N)\ ,
\end{eqnarray}
where $r = \ell/L$ and $2s+1$ is the number of lattice sites.  We make some brief remarks about how these traces were computed below.

Because of the relation $\lambda_k = \frac{1}{4} \coth^2(\epsilon_k/2)$ between the entanglement spectrum and the eigenvalues of $C^2$, we know that the eigenvalues of $C^2$ are bounded below by $1/4$.  The largest even eigenvalue $\lambda_e$ and odd eigenvalue $\lambda_o$ are thus bounded above by
\begin{eqnarray}
\label{evenupper}
\lambda_e &\leq& \tr C_e^2 - \frac{s}{4} \ , \\
\label{oddupper}
\lambda_o &\leq& \tr C_o^2 - \frac{s-1}{4} \ .
\end{eqnarray}

\begin{figure}
 \begin{center}
a) \includegraphics[width=2.5in]{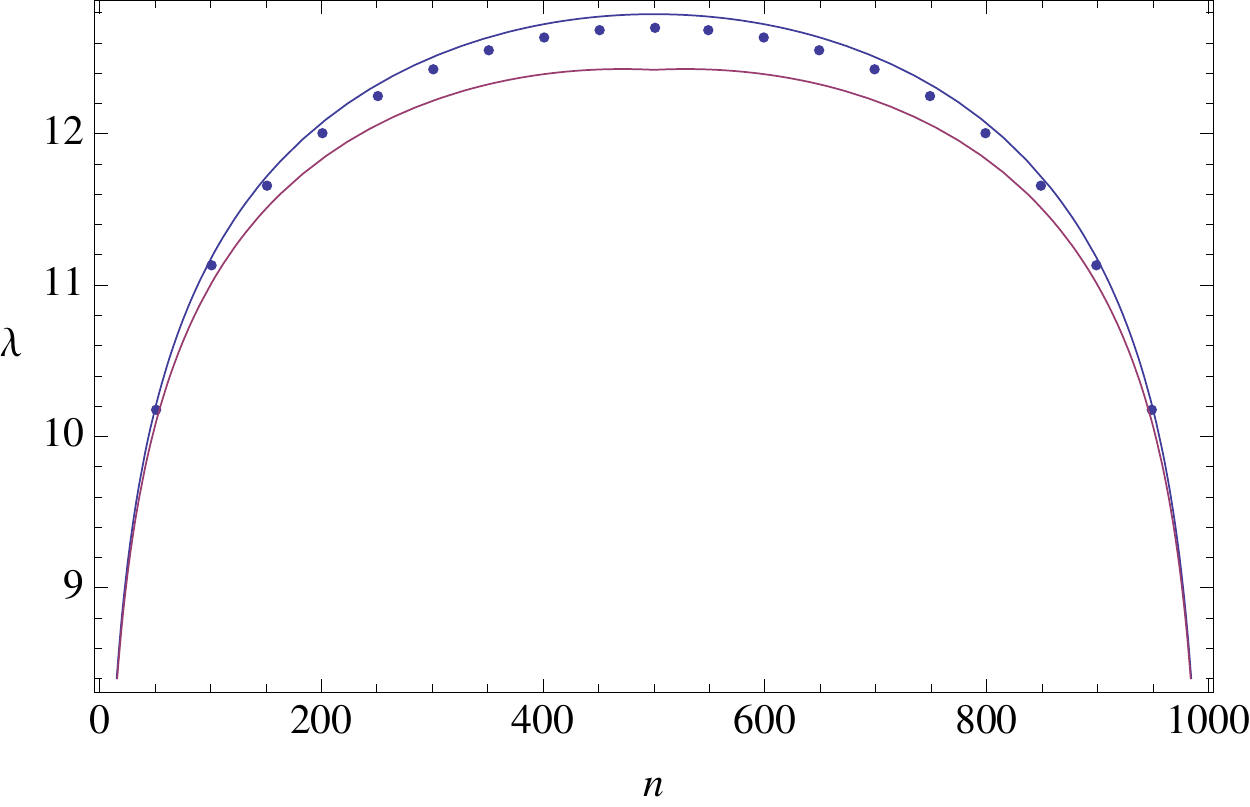} 
b)\includegraphics[width=2.5in]{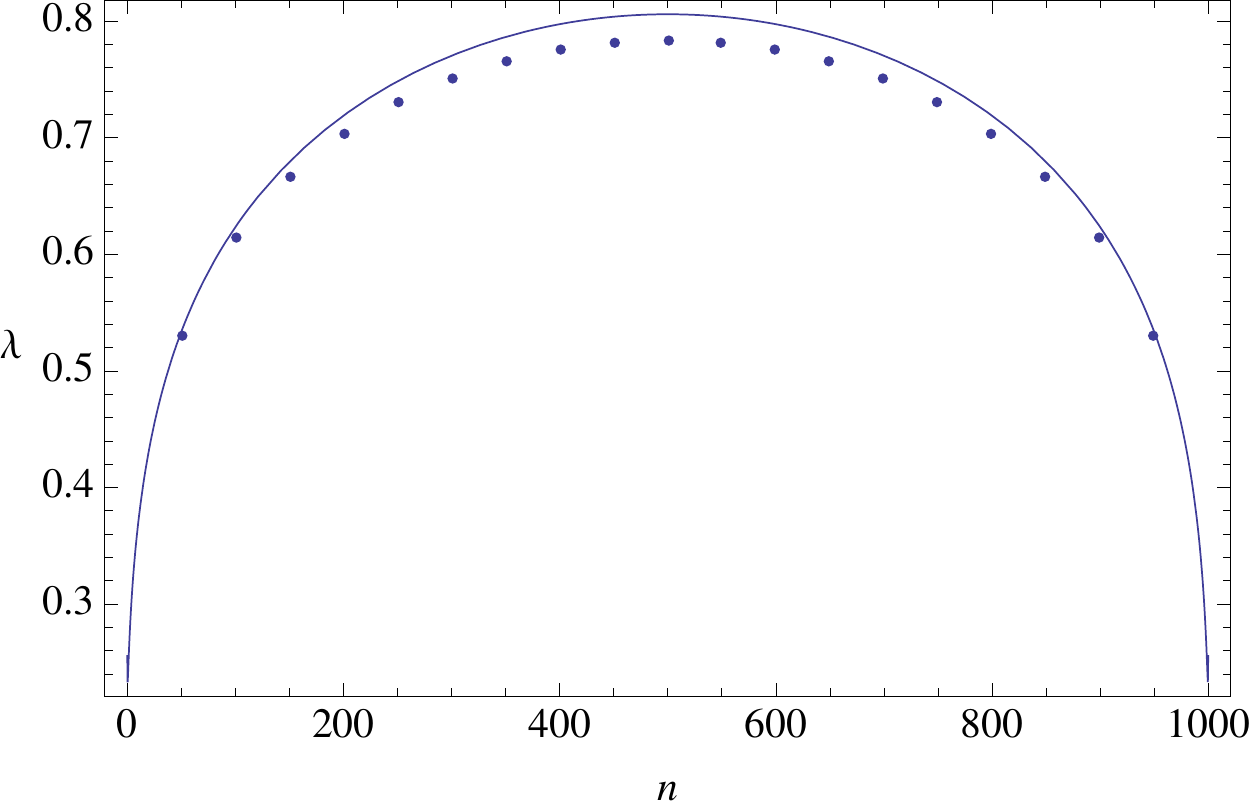}
 \end{center}
 \caption{
The largest (a) and second largest (b) eigenvalue of $C^2$ plotted against the interval length for $mL=1/10$, $T=0$, and $N=1000$.  The points are numerically computed.  The curves above the points are the analytic upper bounds (\ref{evenupper}) and (\ref{oddupper}) computed from the traces.  The solid curve below the points on the left is the lower bound (\ref{lowerbound}) computed from the variational principle.
\label{boundplot}
  }
 \end{figure}

We can also put a lower bound on $\lambda_e$ by using the variational principle and a ``trial wave function''.  In this case, we use a constant trial wave function, $\psi_e = (1, 1, \ldots, 1) / \sqrt{n}$.  The expectation value then provides a lower bound:
    \begin{eqnarray}
    \label{lowerbound}
  \lambda_e &\geq&  \langle \psi_e | C_e^2 | \psi_e \rangle \\
  &=& \frac{1}{2\pi mL} \coth \left(\frac{m}{2T} \right) \left[ \gamma + \ln \left( \frac{4N \sin ( \pi r)}{\pi} \right) \right] 
 + \frac{1}{12}
\nonumber
\\
&&
- \frac{i}{8 \pi^3 r} \left[ \gamma + \ln \left( \frac{4 N \sin(\pi r)}{\pi} \right) \right]
\left[ \operatorname{Li}_2(e^{2 \pi i r}) - \operatorname{Li}_2(e^{-2 \pi i r}) \right]
\nonumber \\
&&
-\frac{r^2}{4} \left[ \frac{1}{3} - \coth^2 \left( \frac{m}{2T} \right) \right] + O(m L,e^{-2\pi/ TL}, \log N /N)
\nonumber  
 \ .
  \end{eqnarray}
  Figure \ref{boundplot} demonstrates that our upper and lower bounds provide relatively good estimates of the two largest eigenvalues at $T=0$.  
  We could  
try to produce an analytic lower bound on $\lambda_o$ by similar methods.  
However, 
simple trial wave functions such as $(\psi_o)_j \sim \sin (\pi j / N)$ or $(\psi_o)_j \sim j$
do not seem to give strong lower bounds numerically and are harder to work with analytically than the constant trial wave function used above in the even case.

The zero mode $a=0$ terms in $\langle \phi \phi \rangle$ and $\langle \pi \pi \rangle$ have a large influence on the structure of these traces in our $m, T \ll 1/L$ limit.  
As these zero modes have even parity, they do not contribute to $C_o^2$.  
For example, note that $\tr C_e^2 = O(1/m L)$ is much larger than $\tr C_o^2 = O(1)$ because
the zero mode $a=0$ term in $\langle \phi \phi \rangle$ 
is $O(1/m L)$ but only contributes to the even sector of $C^2$.   
Also note that only $\tr C_e^2$ depends on $T$.  The reason is that $\coth (\omega_a / 2T )\approx 1$ up to exponentially suppressed terms except when $a=0$.

\begin{figure}
 \begin{center}
a) \includegraphics[width=2.4in]{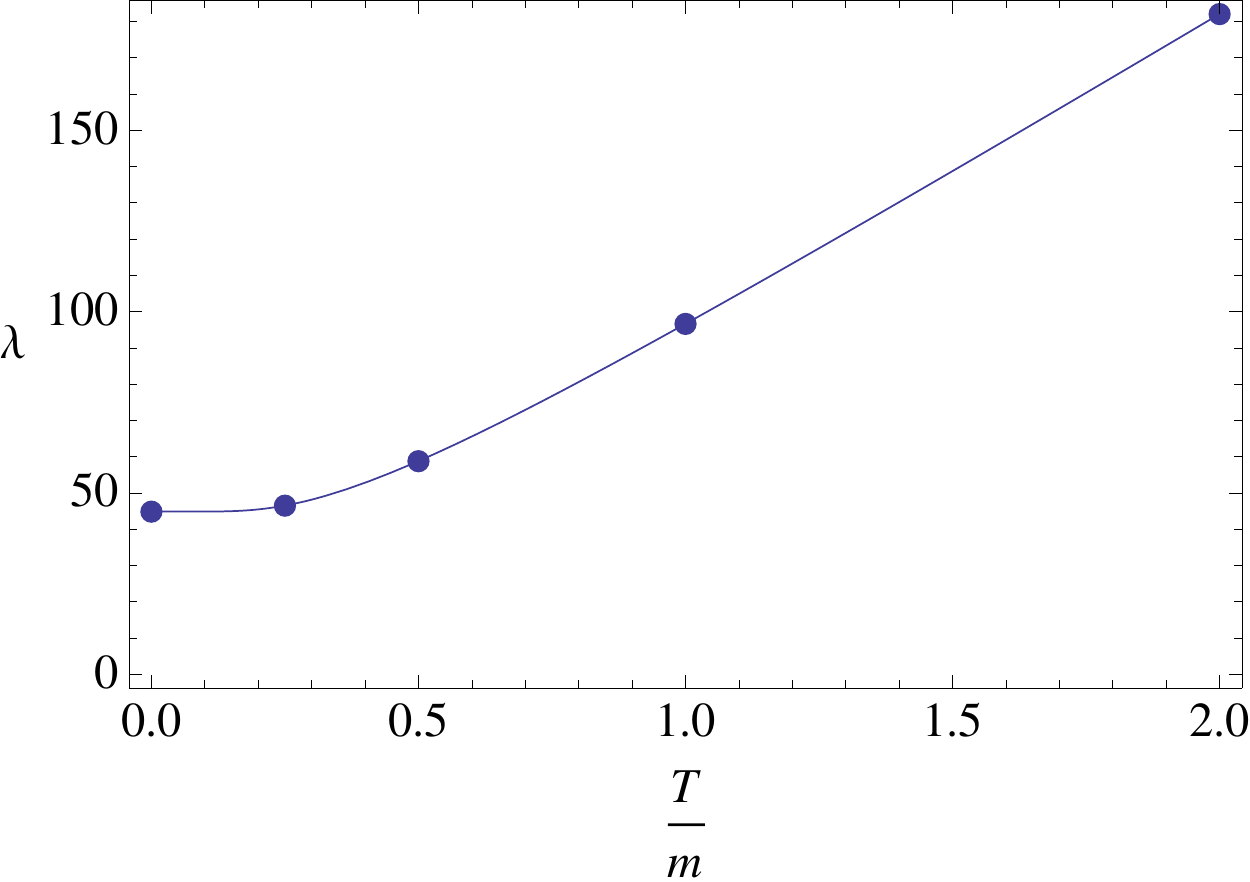}
  b)\includegraphics[width=2.4in]{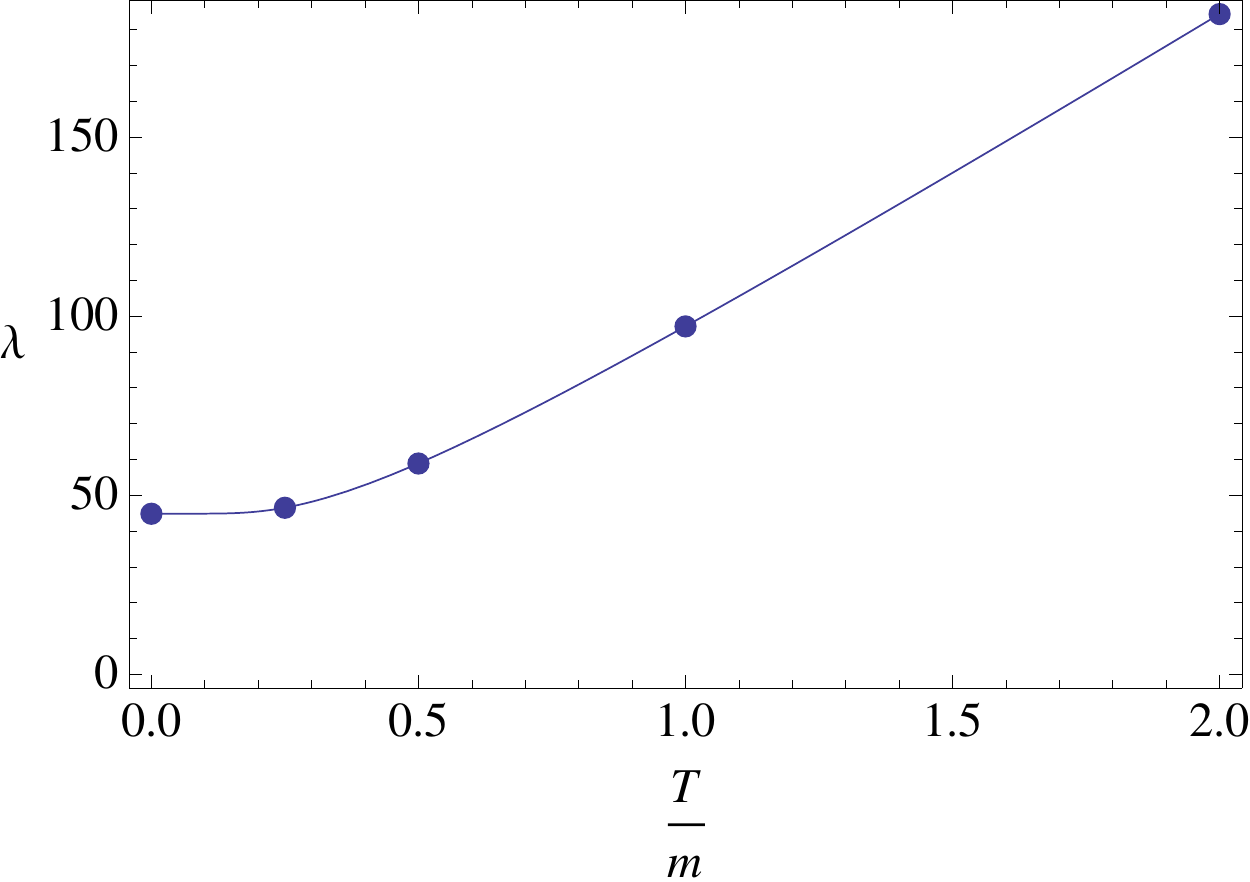}
 \end{center}
 \begin{center}
 c)\includegraphics[width=2.4in]{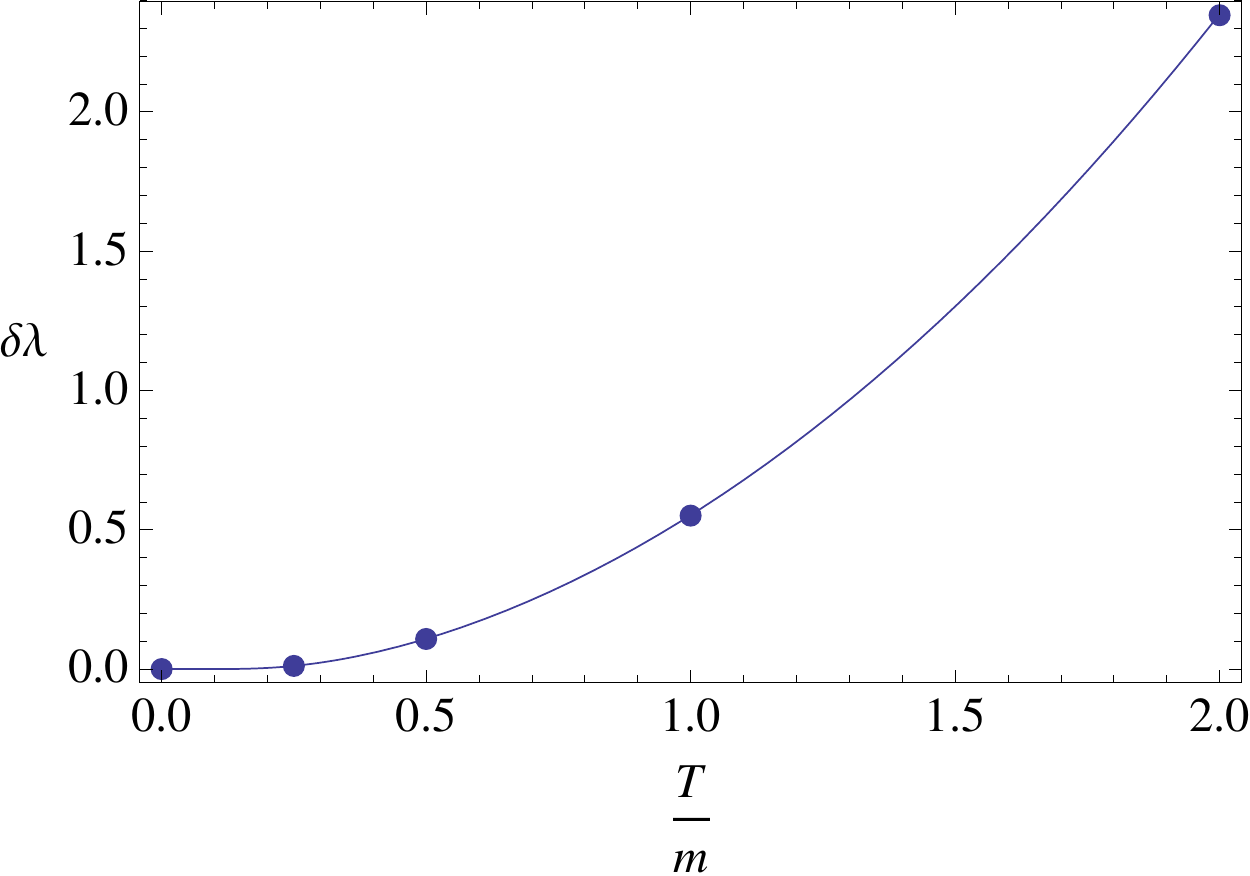}
 \end{center}
 \caption{
The largest eigenvalue of $C^2$ as a function of temperature for $mL = 1/50$: a) $\ell/L=1/5$; b) $\ell/L=4/5$; c) the difference between the two for a lattice with $N=200$.  
The points are numerical while the curve is the upper bound computed from $ \tr C_e^2$.
\label{bound2plot}
  }
 \end{figure}

Another interesting feature of these traces is their behavior under the exchange of the interval $A$ with its complement $B$.  By translation invariance, this exchange can be implemented by sending $r \to 1 -r$.  At $T=0$, both $\tr C_o^2$ and $\tr C_e^2$ are invariant under this transformation.  This invariance is expected in order to guarantee that $S_A = S_B$.  
For $T\neq 0$, the breaking of this symmetry is due entirely to the
$r^2  \csch^2 (m/2T)$ term in $\tr C_e^2$.\
This symmetry breaking term comes from multiplying the $a=0$ zero modes in $\langle \phi \phi \rangle$ and $\langle \pi \pi \rangle$ together.
Figure \ref{bound2plot} demonstrates that $\tr C_e^2$ gives a remarkably good estimate of the temperature dependence of the largest eigenvalues for regions $A$ and $B$, and also for their difference.

 We should say a few words about the lengthy computation performed to obtain (\ref{trCe2}), (\ref{trCo2}), and (\ref{lowerbound}).  
Consider first the $O(1/mL)$ contribution to $\tr C_e^2$:
\be
\label{trCe2initial}
\tr C_e^2 = \frac{1}{2 m L} \coth \left(\frac{m}{2T} \right) f(n,N) + O(m L)^0 + O(e^{-2 \pi / TL}) \ ,
\ee
where
\be
f(n, N) \equiv \frac{1}{N} \sum_{a=1}^N \frac{ \sin^2 (\pi a n / N) }{\sin (\pi a / N)} = \frac{1}{N} \sum_{j=1}^{n} \cot \frac{\pi}{N} (j-1/2 ) \ .
\ee
We want to evaluate this sum in the continuum limit where $n$ and $N$ are both large but $r = n/N$ is held fixed between zero and one.
Replacing the sum over $j$ by an integral introduces unacceptably large errors because of the divergence at $j=1/2$.  Instead, we compute a related integral that does not have this divergence:
\begin{eqnarray}
\label{fnN}
f(n,N) 
&\approx& \int_{1/N}^r \left( \cot \pi (x - 1/2N) - \frac{1}{\pi(x-1/2N)} \right) dx \nonumber \\
&& \; \; \; \; + \frac{1}{N}\sum_{j=1}^n \frac{N}{\pi (j-1/2)}  \\
&=& \frac{1}{\pi} \left[ \ln \left( \frac{4 N \sin (\pi r)}{\pi} \right) + \gamma \right] 
+ O(1 /N^2) \ .
\end{eqnarray}

Calculating the $O(m L)^0$ and $O(mL)$ terms is a more complicated enterprise.  
 As mentioned already above, one contribution to $C_e^2$ comes from multiplying the zero modes in $\langle \phi \phi \rangle$ and $\langle \pi \pi \rangle$ together and yields $r^2 \csch^2(m/2T)$. 
  The remaining order one pieces can be computed from the matrix $C_e^2$ with the zero modes removed in the limit $m=0=T$:
 \begin{eqnarray}
 \label{CeOzero}
 (\tilde C_e^2)_{jk} &=& \frac{1}{4N^2} \sum_{a,b=1}^{N-1} \sum_{l=-s}^s
 \frac{\sin \frac{\pi a}{ N}}{\sin \frac{\pi b}{ N}} \cos \frac{2\pi l a}{N} \cos \frac{2 \pi l b}{N} \cos \frac{2 \pi j a}{N} \cos \frac{2 \pi k b}{N} \ .
 \end{eqnarray}
 Similarly, the $O(m L)^0$ contribution to $\tr C_o^2$ can be calculated from the
$m=T=0$ limit of the matrix $C_o$:
 \begin{eqnarray}
 \label{CoOzero}
 (\tilde C_o^2)_{jk} &=& \frac{1}{N^2} \sum_{a,b=1}^{N-1} \sum_{l=1}^s
 \frac{\sin \frac{\pi a}{ N}}{\sin \frac{\pi b}{ N}} \sin \frac{2\pi l a}{N} \sin \frac{2 \pi l b}{N} \sin \frac{2 \pi j a}{N} \sin \frac{2 \pi k b}{N} \ .
 \end{eqnarray}
The $O(mL)$ term of $(C_e^2)_{jk}$ comes from zero modes pieces of $C^2$ where either $a=0$ in the $\langle \phi \phi \rangle$ sum or $a=0$ in the $\langle \pi \pi \rangle$ sum:
\be
\label{CeOone}
\frac{3 m L}{16 N^3} \coth \left( \frac{m}{2T} \right) \sum_{b=1}^{N-1} \sum_{l=-s}^s \frac{\cos \frac{2 \pi l b}{N} \cos \frac{2 \pi k b}{N}}{\sin \frac{\pi b}{N}} \ .
\ee
(For $(C_e)_{jk}$, the indices have the range $-s \leq j,k \leq s$, while for $(C_o)_{jk}$, we restrict to $1 \leq j,k \leq s$.)
In the appendix, 
we describe how to perform the sums (\ref{CeOzero}), (\ref{CoOzero}), and (\ref{CeOone}) along with (\ref{lowerbound}) in the the large $N$ limit with $s/N$ held fixed.

\section{Raising the Temperature}

We present three arguments that the entanglement entropy depends exponentially on the ratio $m/T$ in the limit $T \ll m$.  The first argument is heuristic and relies on the structure of the matrix $C^2$.  The second argument is based on our earlier calculation of $ \tr C_e^2$.  
The third argument is based on numerical evidence.
We would like to show two things.  The first is that for a fixed interval $A$, 
\be
S(T) - S(0) \sim e^{-m/T} \ .
\label{STone}
\ee
The second is that for two complementary intervals $A$ and $\bar A = B$,
\be
S_A - S_B \sim e^{-m/T} \ .
\label{STtwo}
\ee

The first argument relies on the fact that the temperature dependence of $C^2$ comes entirely from the factors of 
$\coth(\omega_a / 2  T)$ in $\langle \phi \phi \rangle$ and $\langle \pi \pi \rangle$.  The frequency $\omega_a$ is bounded below by $m$.  Thus we conclude that
\be
\coth\left(\frac{\omega_a}{2  T} \right) \leq \coth \left( \frac{m}{2T} \right ) = 1 + 2 e^{-m/T} + O(e^{-2m/T}) \ .
\ee
In other words, the matrix $C$ has a low temperature expansion of the form
\be
C(T) = C(0) + e^{-m/T}  \delta C + \ldots
\ee
where the ellipsis denotes terms that are more exponentially suppressed.
Now if $C(T)$ has such an expansion, then the eigenvalues $\nu_k(T) = \nu_k(0) + e^{-m/T} \delta \nu_k +\ldots$ will as well.  
Assuming $\nu_k(0) -1/2 \gg e^{-m/T}$, expanding eq.\ (\ref{SEfromC2}) in the small $T$ limit, one concludes that the entanglement entropy for a single interval shifts by an amount
\be
\delta S = 2 \sum_k  [ \ln (\nu_k(0) + 1/2) - \ln (\nu_k(0) - 1/2) ]  \delta \nu_k e^{-m/T} + \ldots\ ,
\ee
implying the scaling  (\ref{STone}).
Assuming $\delta S$ is different for an interval and its complement, one also concludes the scaling (\ref{STtwo}).

While, the numerical evidence we present below suggests both scalings (\ref{STone}) and (\ref{STtwo}) are correct, there are some loop holes in our argument.  An obvious problem is that the $e^{-m/T}$ term in the small $T$ expansion may vanish; the temperature dependence may be of the form $e^{M/T}$ for some $M>m$.   
A more subtle loop hole involves the fact that many of the $\nu_k(0)$ are close to $1/2$.  In this case, the correction to the entanglement entropy $\delta S$ can scale as $(m/T) e^{-m/T}$ instead of just $e^{-m/T}$.  Numerically, we see no evidence for this behavior.
Instead, in these cases we find that the logarithmic enhancement is not enough to make up for the smallness of $\delta \nu_k$; these eigenvalues contribute negligibly to the entanglement entropy.
%

The second argument for the scalings (\ref{STone}) and (\ref{STtwo}) is based on using 
$\tr C_e^2$ as an estimate of the largest eigenvalue $\lambda_e$.
Using $\tr C_e^2$, we estimate the contribution of $\lambda_e$ to $S$ and infer the scalings from this contribution.
The temperature dependence of a single interval 
comes principally from the leading $\coth(m/2T) /m L$ term in (\ref{trCe2}).  One finds agreement with (\ref{STone}):
\be
\left. [S(T) - S(0)]\right|_{\lambda_e} \sim e^{-m/T} \ .
\ee
Next we consider the entanglement difference $S_B - S_A$.
This type of temperature dependence comes from the $r^2 \csch^2(m/2T)$ piece of (\ref{trCe2}).  One finds agreement with (\ref{STtwo}):
\be
\left. [S_B(T) - S_A(T)] \right|_{\lambda_{e,A}, \lambda_{e,B}} \sim \frac{\pi}{2} \frac{m L}{\log N}\left(1 - 2r \right)e^{-m/T} \ .
\label{dSapprox}
\ee

\begin{figure}
 \begin{center}
  \includegraphics[width=3.2in]{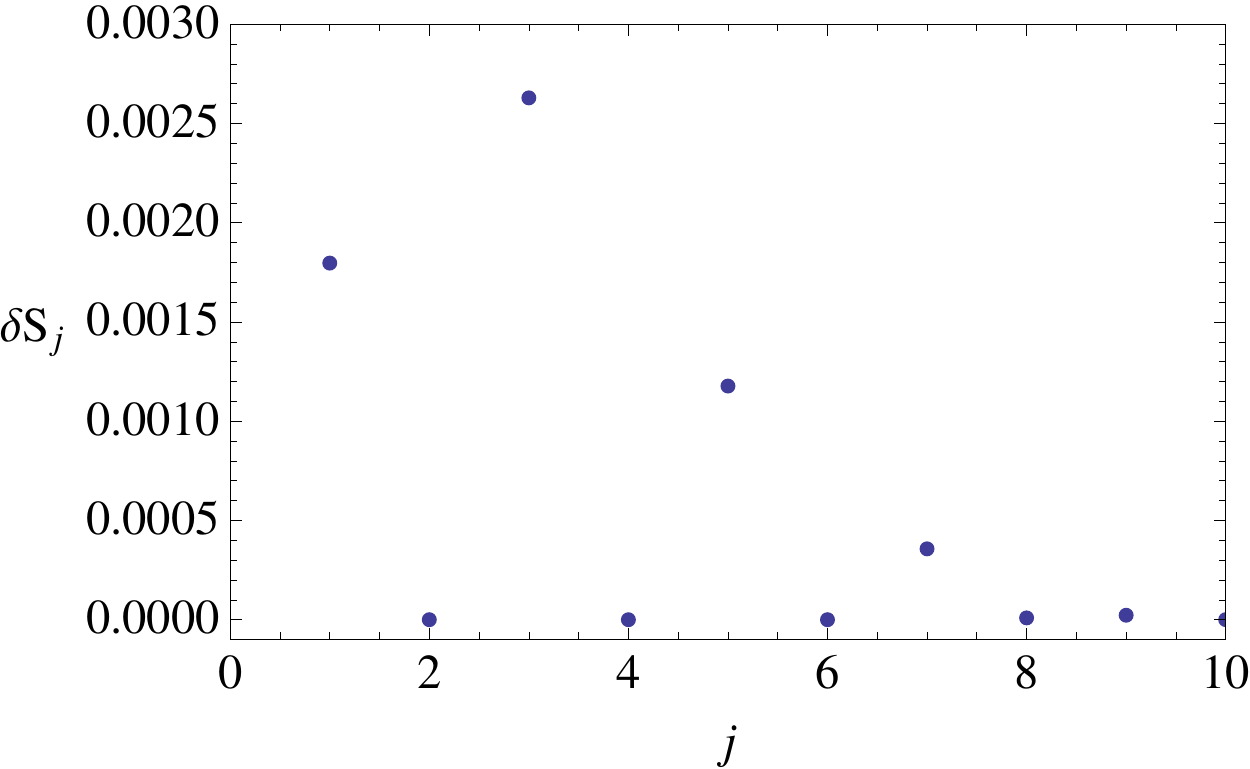}
  \end{center}
  \caption{
The contribution to $\delta S = S_A - S_B$ for the ten largest pairs of eigenvalues $\lambda_{j,A}$ and $\lambda_{j,B}$, arranged from largest to smallest.  In this plot $mL = 0.02$, $m/T = 1$ and $\ell/L = 1/5$ for region $B$.  Note that the odd parity eigenvalues do not contribute.  ($N=200$ was used for this plot.)
}
\label{evalplot}
 \end{figure}

We should emphasize that using $\tr C_e^2$ and the largest eigenvalue $\lambda_e$ to estimate the temperature scalings is flawed.
An obvious limitation is that we only have a result for $\tr C_e^2$ in the limit $mL \ll 1$ while we expect the temperature scalings to hold more generally.
A less obvious limitation is that
despite the fact that $\lambda_e$ is much larger than the other eigenvalues in the small mass limit, the logarithms in (\ref{SEfromC2}) play a democratizing role and let smaller eigenvalues contribute substantially to the entanglement entropy.
For example, in this small mass limit numerical analysis shows that the dominant contribution to $S_A-S_B$ comes from the second largest even eigenvalue (see figure \ref{evalplot}).

\begin{figure}
 \begin{center}
a)  \includegraphics[width=2.7in]{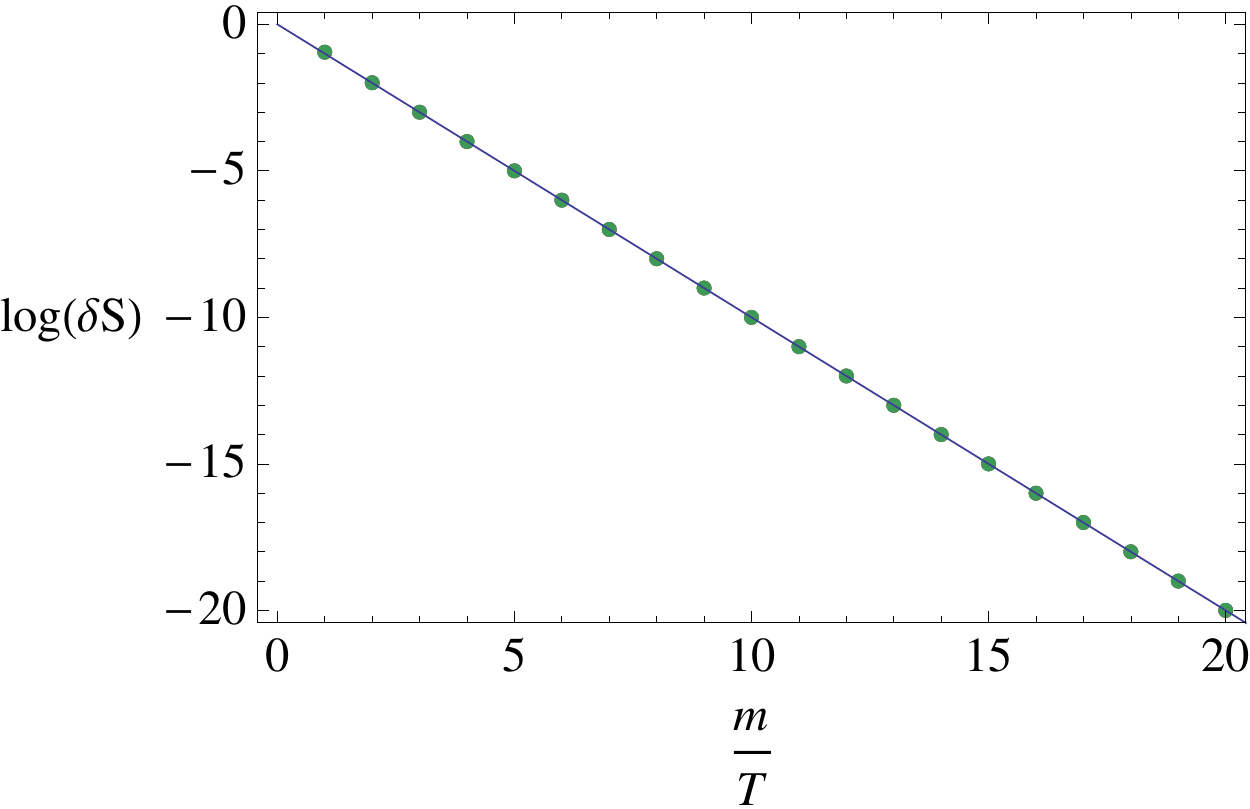}
  b) \includegraphics[width=2.7in]{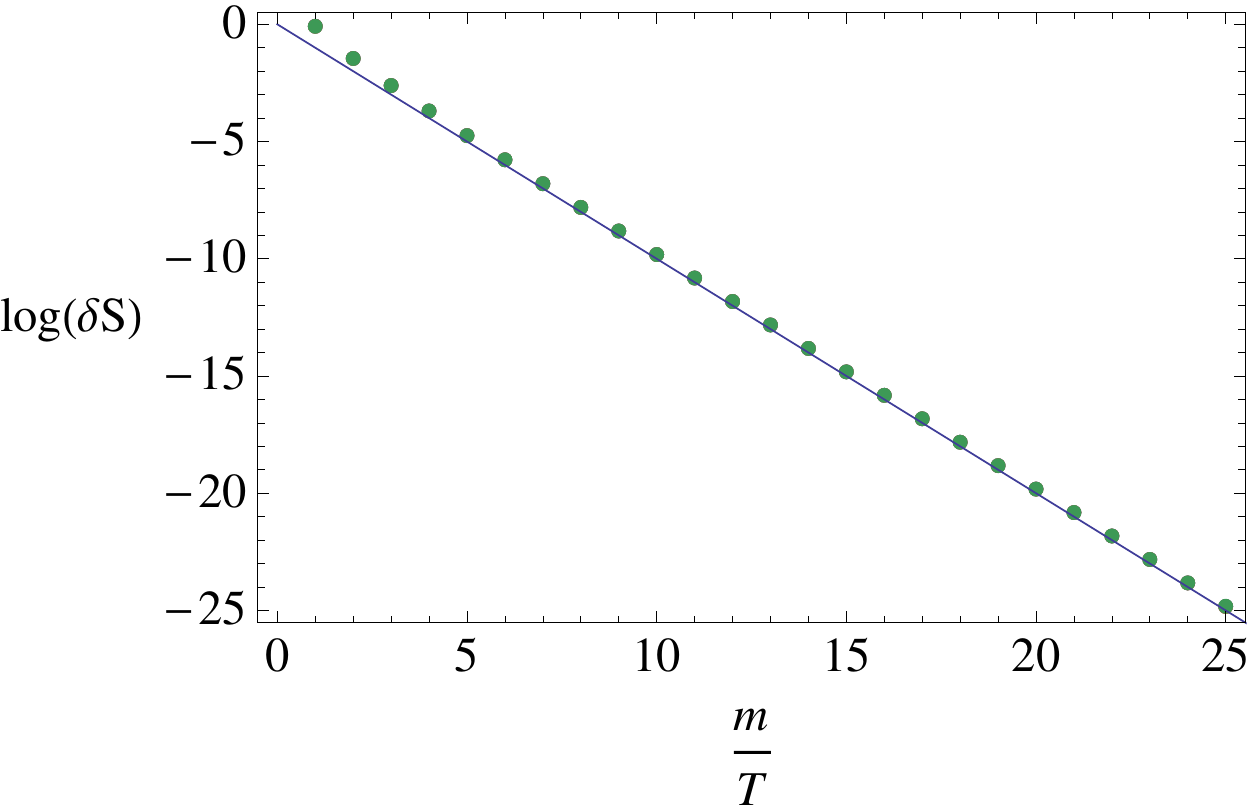}
 \end{center}
  \caption{
A log plot of the entanglement entropy $\delta S = S(T) - S(0)$ vs.\ $m/T$ with an interval size $\ell/L = 3/10$.  The points are numerically computed, and the line $\log(\delta S) = -m/T$ is a guide to the eye:
a) $m L = 5 \times 10^{-3}$;   b) $m L = 5$.
(For both plots, the points were computed with $N=50$, 100, 200, and 400. 
The data points for different values of $N$ all lie roughly on top of each other.)}
\label{SEAinterval}
 \end{figure}

\begin{figure}
 \begin{center}
 a) \includegraphics[width=2.8in]{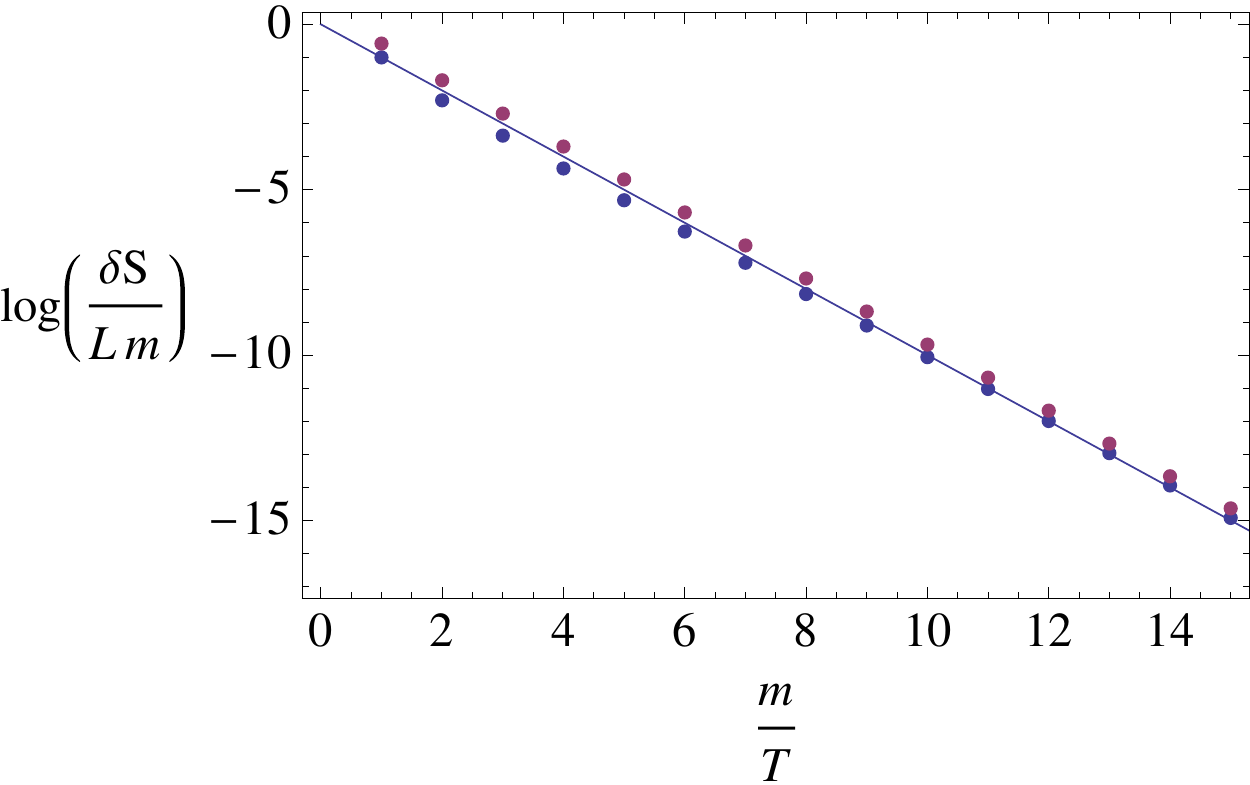}
 b)
 \includegraphics[width=2.6in]{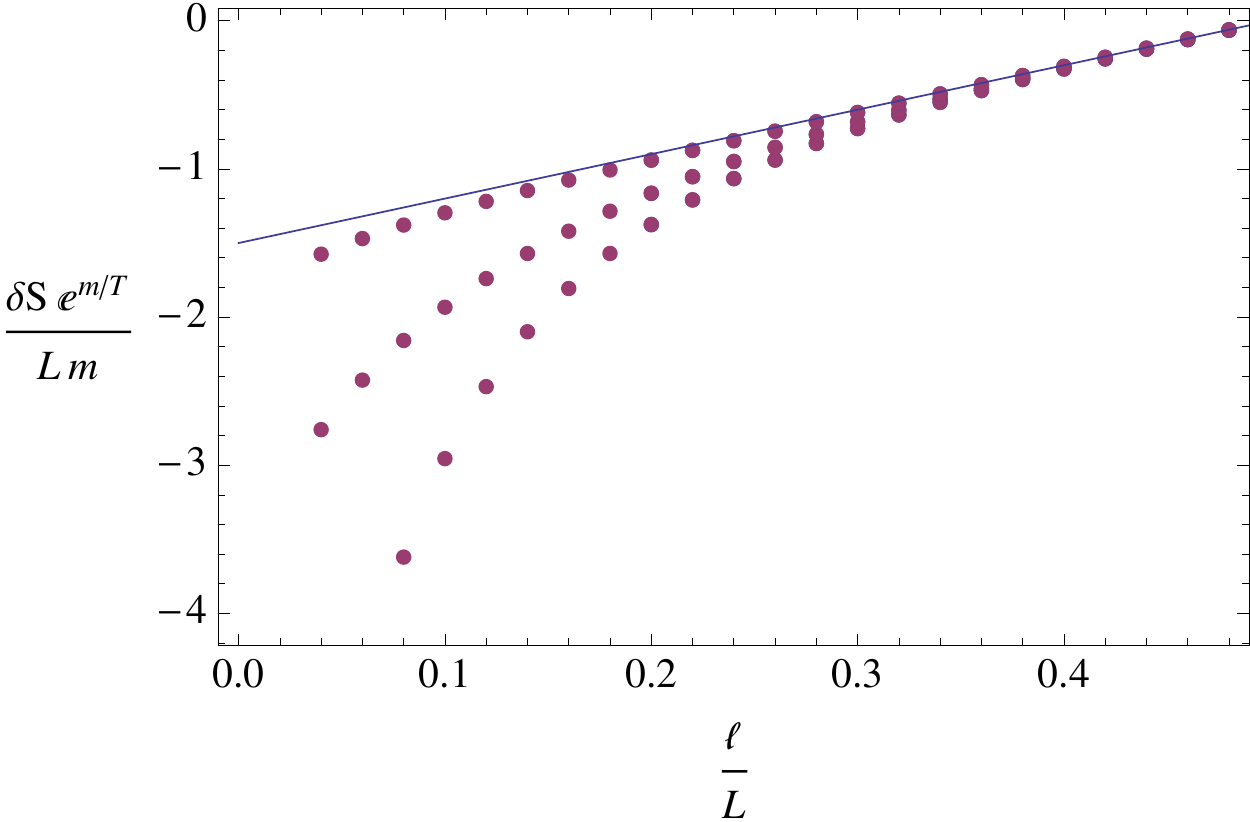}
   \end{center}
  \caption{
a)
A log plot of the entanglement entropy difference $\delta S = S_A - S_B$ vs.\ $m/T$ for $m L = 5$ and $5 \times 10^{-3}$, and an interval B of size $\ell/L= 1/5$.  At fixed $m/T$, the larger mass points lie below the smaller ones.  The line $\log(\delta S / m L) = -m/T$ is a guide to the eye.  (The lattice was taken to have size $N=200$, but there is no noticeable difference between this graph and a graph with $N=100$.)
b)
The entanglement entropy difference $\delta S$ vs.\ $\ell/L$ for (from bottom to top) $mL = 5 \times 10^{-3}$, $2$, and $5$.  The mass to temperature ratio is $m/T = 10$.  The line $e^{m/T} \delta S  / m L = 3m / T - 3/2$ is a guide to the eye.
(The lattice was taken to have $N=400$, but there is no difference between this graph and a graph with $N=200$.)
}
\label{Sdiffplot}
 \end{figure}

Our most convincing evidence for the scalings (\ref{STone}) and (\ref{STtwo}) is numerical and is presented in figures \ref{SEAinterval} and \ref{Sdiffplot}.  
Figure \ref{SEAinterval} demonstrates unambiguous evidence for (\ref{STone}), not only for $mL \ll 1$ but also for $mL > 1$.  
Figure \ref{Sdiffplot}a displays unambiguous evidence for (\ref{STtwo}), again both for small and large values of $mL$.
More ambitiously, we can try to investigate numerically whether the $m L (1-2r) / \log N$ behavior of eq.\ (\ref{dSapprox}) is correct as well.
Figure \ref{Sdiffplot}a provides evidence for the $mL$ scaling.  Figure \ref{Sdiffplot}b provides some limited evidence for the $1-2r$ behavior for large values of $mL$ and for intervals with $r \sim 1/2$.  However, we find no evidence for the $\log N$ behavior of (\ref{dSapprox}).

\section{Discussion}

As mentioned in the introduction, the original motivation for this paper came from the Ryu-Takayanagi proposal \cite{Ryu:2006bv} for computing the entanglement entropy of field theories with holographic dual classical gravity descriptions.  
In their proposal, the field theory lives on the boundary of the space-time in the dual description.  Let $C$ be the curve that separates region $A$ from region $B$ in the field theory.  Let $C$ also be the boundary of a minimal surface $M$ that falls into the space-time.  The proposal is that the entanglement entropy is proportional to the area of $M$:
\be
S_A = \frac{\operatorname{Area}(M)}{4 G_N} \ ,
\ee
where $G_N$ is Newton's constant.  Assuming a unique such $M$, the entanglement entropy of a region and its complement are always equal, $S_A = S_B$.

\begin{figure}
\begin{center}
\includegraphics[width=4in]{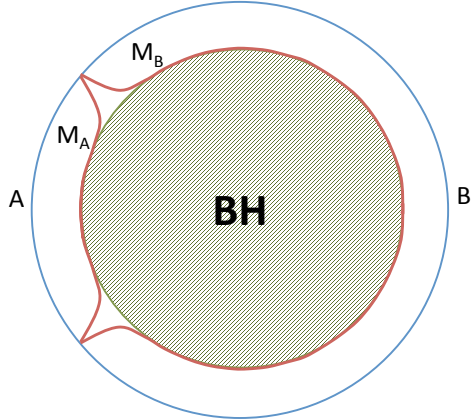}
\end{center}
\caption{
The two minimal surfaces $M_A$ and $M_B$ corresponding to a region $A$ and its complement $B$ when the dual space time contains a black hole (BH).
\label{RT}
}
\end{figure}

When the space-time contains a black hole, Ryu-Takayanagi modified their proposal to account for the existence of two minimal surfaces $M_A$ and $M_B$.  
The entanglement entropy for $A$ must be computed from the surface $M_A$ that is deformable into $A$.  Correspondingly, for region $B$, we must use $M_B$.  For 
large black holes, $\operatorname{Area}(M_A) - \operatorname{Area}(M_B)$ will come 
mostly from the differing amount of black hole horizon area that the two surfaces wrap (see figure \ref{RT}).  
The Hawking temperature of the black hole corresponds to the temperature of the field theory, and thus this modification of the proposal provides a way for $S_A - S_B$ to be nonzero for certain thermal field theories.  

However, there are instances where field theories at $T>0$ have dual gravity descriptions without a black hole.  A classic example is the large $N$, strong coupling limit of maximally supersymmetric $SU(N)$ Yang-Mills theory on $S^3 \times S^1$ \cite{Witten:1998zw}.  At temperatures small compared to the inverse radius of the $S^3$, the dual description is thermal $AdS_5 \times S^5$.  At a critical temperature $T_c$, the gravity description undergoes a first order Hawking-Page phase transition to a state with a large black hole.  For the field theory, this transition is understood as a deconfinement phase transition.

On the one hand, their proposal implies that the entanglement entropy will serve as an order parameter for the phase transition:
for $T< T_c$, $S_A - S_B =0$, while for $T>T_c$, $S_A - S_B \neq 0$.    
On the other, at any finite $N$, we have a system at finite volume for which there can be no phase transitions.  The transition from 
$S_A  = S_B$ at $T=0$ to $S_A \neq S_B$ at $T>T_c$ must be smooth.
We conclude that the Ryu-Takayanagi formula is only valid in the strict large $N$ limit, 
but it would be nice to understand the form of the $1/N$ corrections.  In principle, one should be able to compute the entanglement entropy for maximally supersymmetric Yang-Mills at weak coupling.  In practice, such a computation 
is substantially more difficult, and
we instead considered a 1+1 dimensional massive scalar field on a circle at $T>0$.  
Morally, the regime $T<m$ should correspond to the confining regime of the Yang-Mills theory where the fields get a mass through their coupling to the curvature of the $S^3$.
For our scalar field, we argued that in the regime $T \ll m$, the entanglement entropy difference scales as
\[
S_B - S_A \sim e^{-m/T} \ .
\]
We conjecture that this type of scaling should be a generic feature of all gapped systems.

\subsection*{Acknowledgements}

We would like to thank A.~Abanov, K.~Balasubramanian, P.~Gao, G.~Giecold, D.~Gulotta, T.~Nishioka, and T.-C. Wei for discussion.  This work was supported in part by the National Science Foundation under Grants No. PHY-0844827 and PHY-0756966.  C.~H. also thanks the Sloan Foundation for partial support.

\appendix

\section{Computing Traces in the Continuum Limit}

The basic technique used in computing $\tr C_e^2$, $\tr C_o^2$ and $\langle \psi | C_e^2 | \psi \rangle$ in the continuum limit $N \to \infty$ with $n/N$ held fixed was to replace sums with integrals.  However, there are three wrinkles in this procedure, two of which have already been hinted at in the text.  
The first is that we were not able to perform the integrals obtained by taking the continuum limit of the mode sums over $a$ and $b$.  Thus, we first performed the mode sums over $a$ and $b$ explicitly yielding  sums over cotangents.
For example, performing the mode sums for eq.\ (\ref{CeOzero}) yields,
\begin{eqnarray}
 \label{Ceabsum}
8 N^2  (\tilde C_e^2)_{jk} &=& \left[ \cot \frac{\pi}{N} (s+j+1/2) + \cot \frac{\pi}{N}(s-j+1/2) \right]  \times \nonumber \\
&&
\left[ \sum_{b=1}^{N-1} \csc \frac{\pi b}{N} - 2 \sum_{l=1}^s \cot \frac{\pi}{N} (l-1/2) \right. \nonumber \\
&& \left. - \sum_{l=1}^{|k|} \left( \cot \frac{\pi}{N}(s+l-1/2) - \cot \frac{\pi}{N} (s-l+1/2) \right) \right] \nonumber \\
&& + \sum_{l=1}^s \left[ \cot \frac{\pi}{N} (l+j-1/2) + \cot \frac{\pi}{N}(l-j-1/2) \right] \times \nonumber \\
&& \left[ \cot \frac{\pi}{N}(l+k-1/2) + \cot \frac{\pi}{N} (l-k-1/2) \right] \ ,
 \end{eqnarray}
while performing the mode sums for eq.\ (\ref{CoOzero}) gives
\begin{eqnarray}
\label{Coabsum}
(\tilde C_o^2)_{jk} &=& \frac{1}{4 N^2}\sum_{l=1}^s \left[ 
\cot \frac{\pi}{N}(j+l-1/2)  - \cot \frac{\pi}{N} (j+s+1/2) +\right. \nonumber \\
&&
\hspace{20mm}
\left.  +\cot \frac{\pi}{N} (j-l+1/2) - \cot \frac{\pi}{N}(j-s-1/2) \right] \times \nonumber \\
&&
\hspace{20mm}
\times \left[ \cot \frac{\pi}{N}(k+l-1/2)  + \cot \frac{\pi}{N} (k-l+1/2) \right] \ .
\end{eqnarray}

The second wrinkle is that naive integral approximations of the cotangent sums often include singular regions.  Our strategy in this case was to add and subtract a sum that we could perform analytically but whose integral approximation had the same singular region.  This procedure was already used in the text to perform the sum (\ref{fnN}).
The third wrinkle is that the integral approximations of the cotangent sums were often difficult to perform.  Changing variables and using discrete symmetries reduced the integrals to known results in most cases.  However, in two cases, we had to perform an integral we could not find in the books.

Let us first sketch the computation of $\tr C_o^2$, i.e.\ the trace of (\ref{Coabsum}).  Several of the terms in the sum have the structure
\be
I_{\pm \pm} = \frac{1}{N^2}\sum_{k,j=1}^s \cot \frac{\pi}{N} (k\pm( j-1/2)) \cot \frac{\pi}{N} (k\pm ( j-1/2)) \ .
\ee
To perform these sums, we make the change of variables $x=k+j$ and $y = k-j$.  Using the same technique in eq.\ (\ref{fnN}) to regularize the singular regions of the integral approximations, one straightforwardly finds
\begin{eqnarray}
I_{++} &=& - \frac{1}{4} - \frac{s^2}{N^2} + \frac{1}{\pi^2} \left[ \ln \frac{2 N \tan (\pi s / N)}{\pi}  + 1 + \gamma \right] + O(1/N) \ , \\
I_{--} &=& s - \frac{s^2}{N^2} - \frac{2}{\pi^2} \left[ \ln \frac{4 N \sin(\pi s/N)}{\pi} + 1 + \gamma \right] + O(1/N^2) \ , \\
I_{+-}=I_{-+} &=& \frac{1}{8} + O(\log N / N ) \ .
\end{eqnarray}
An intermediate result necessary for the computation of $I_{+-}$ is 
\be
\frac{1}{N} \sum_{k=1}^n (-1)^k \cot \frac{\pi}{N} (k-1/2) = - \frac{1}{2} + \frac{(-1)^n}{2N} \cot \left(\frac{\pi n}{N} \right) + O(1/N^3) \ .
\ee
The remaining pieces of $\tr C_o^2$ can be rearranged in the following way
 \begin{eqnarray}
\lefteqn{ 2 \sum_{k,j=1}^s \left( 
  \cot \frac{\pi}{N}(k+j-1/2) + \cot \frac{\pi}{N}(k-j+1/2) 
  \right) \times}
 \nonumber \\
  \lefteqn{
  \times
   \left(
  \cot \frac{\pi}{N} (k-s-1/2) + \cot \frac{\pi}{N}(k+s+1/2) 
  \right)}
 \nonumber \\
  &=&
- \left( \sum_{y=0}^{2s}  \cot \frac{\pi}{N} (y+1/2) \right)^2- \sum_{y=0}^{2s} \cot^2 \frac{\pi}{N}(y+1/2) 
\nonumber \\
&&
+ 2 \sum_{y=0}^{2s} \sum_{x=2s-y}^{2s}
\cot \frac{\pi}{N} (y+1/2) \cot \frac{\pi}{N}(x+1/2) \ .
\label{mess}
\end{eqnarray}
The first sum on the r.h.s.\ of eq.\ (\ref{mess}) we performed in (\ref{fnN}).  The second sum can be performed using the same techniques:
\be
\frac{1}{N^2} \sum_{k=1}^{n} \cot^2 \frac{\pi}{N} (k-1/2) = \frac{1}{2} + O(1/N) \ .
\ee
The third sum requires more work and reduces to one of the two integrals we could not find in tables.
Up to $\log N/ N$ corrections, we may replace the third sum by the following integral:
\be
I(b) \equiv
\int_0^1 \int_{1-x}^1 \cot(b x) \cot(b y) \, dy \, dx \ ,
\ee
where 
\be
\frac{1}{N^2} \sum_{y=0}^{2s} \sum_{x=2s-y}^{2s}
\cot \frac{\pi}{N} (y+1/2) \cot \frac{\pi}{N}(x+1/2)
= \left(\frac{2 s}{N} \right)^2 I \left( \frac{ 2 \pi s }{ N}\right) + O(\log N/ N) \ .
\ee
The integral over $dy$ is trivial:
\be
I(b) = \frac{1}{b} \int_0^1 \cot(bx) \log \frac{\sin(b)}{\sin(b(1-x))} dx \ .
\ee
We find that $I'(b)  b + 2 I(b) = -1$ and that in the small $b$ limit $I(b) = \pi^2 / 6 b^2 + O(1)$.  From these two facts, we deduce that\footnote{
We would like Dan Gulotta for showing us how to perform this integral and also the integral (\ref{Jb}).
}
\be
I(b) = \frac{\pi^2}{6b^2} - \frac{1}{2} \ .
\ee

The quanties $\tr C_e^2$ and $\langle \psi | C_e^2 | \psi \rangle$ may be computed in an analogous way.  
As can be seen in eq.\ (\ref{Ceabsum}), there was one mode sum we were forced to do in the continuum limit:
\be
\frac{1}{N} \sum_{b=1}^{N-1} \csc \frac{\pi b}{N} = \frac{2}{\pi} \left( \gamma + \ln \frac{2N}{\pi} \right) + O(1/N^2) \ .
\ee
All of the other mode sums we were able to perform explicitly.  The remaining sums over cotangents are similar to cases treated above.
We spare the reader almost all of the remaining details. 
In the computation of $\langle \psi | C_e^2 | \psi \rangle$, we came across a second novel integral:
\be
\label{Jb}
J(b) = \int_0^1 \left[ \log \frac{\sin(b(1+x))}{\sin(b(1-x))} \right]^2 dx \ .
\ee
Similar to the strategy in computing $I(b)$, we find that $J''(b) + 2 J'(b)/b = -8$ and that in the small $b$ limit
$J(b) = \pi^2 / 3 + O(b^2)$.  Thus we deduce that
\be
J(b) = \frac{1}{3} ( \pi^2 - 4 b^2) \ .
\ee

\end{document}